\documentclass{acm_proc_article-sp}
\usepackage{graphicx}
\usepackage{float}
\usepackage{afterpage}
\usepackage{wrapfig}

\begin{document}

\conferenceinfo{}{Bloomberg Data for Good Exchange 2016, NY, USA}

\title{Can Big Media Data Revolutionize Gun Violence Prevention?}

\numberofauthors{5}
\author{
\alignauthor
John W. Ayers\\
       \affaddr{San Diego State University}\\
       \affaddr{San Diego, CA, USA}\\
       \email{ayers.john.w@gmail.com}
\alignauthor
Benjamin M. Althouse\\
       \affaddr{Santa Fe Institute}\\
       \affaddr{Santa Fe, NM}\\
       \email{althouse@santafe.edu}
\and  
\alignauthor 
Eric C. Leas\\
       \affaddr{UC San Diego}\\
       \affaddr{La Jolla, CA, USA}\\
       \email{eleas@ucsd.edu}  
\alignauthor   
Ted Alcorn \\
       \affaddr{Everytown for Gun Safety}\\
       \affaddr{New York, New York}\\
       \email{talcorn@everytown.org}         
\alignauthor 
Mark Dredze\\
       \affaddr{Johns Hopkins University}\\
       \affaddr{Baltimore, MD, USA}\\
       \email{mdredze@cs.jhu.edu}
}

\maketitle

\begin{abstract}
The scientific method drives improvements in public health, but a strategy of obstructionism has impeded scientists from gathering even a minimal amount of information to address America's gun violence epidemic. We argue that in spite of a lack of federal investment, large amounts of publicly available data offer scientists an opportunity to measure a range of firearm-related behaviors. Given the diversity of available data -- including news coverage, social media, web forums, online advertisements, and Internet searches (to name a few) -- there are  ample opportunities for scientists to study everything from trends in particular types of gun violence to gun-related behaviors (such as purchases and safety practices) to public understanding of and sentiment towards various gun violence reduction measures. Science has been sidelined in the gun violence debate for too long. Scientists must tap the big media datastream and help resolve this crisis.  
\end{abstract}

\begin{figure*}[t]
\centering
\includegraphics[width=.99\textwidth]{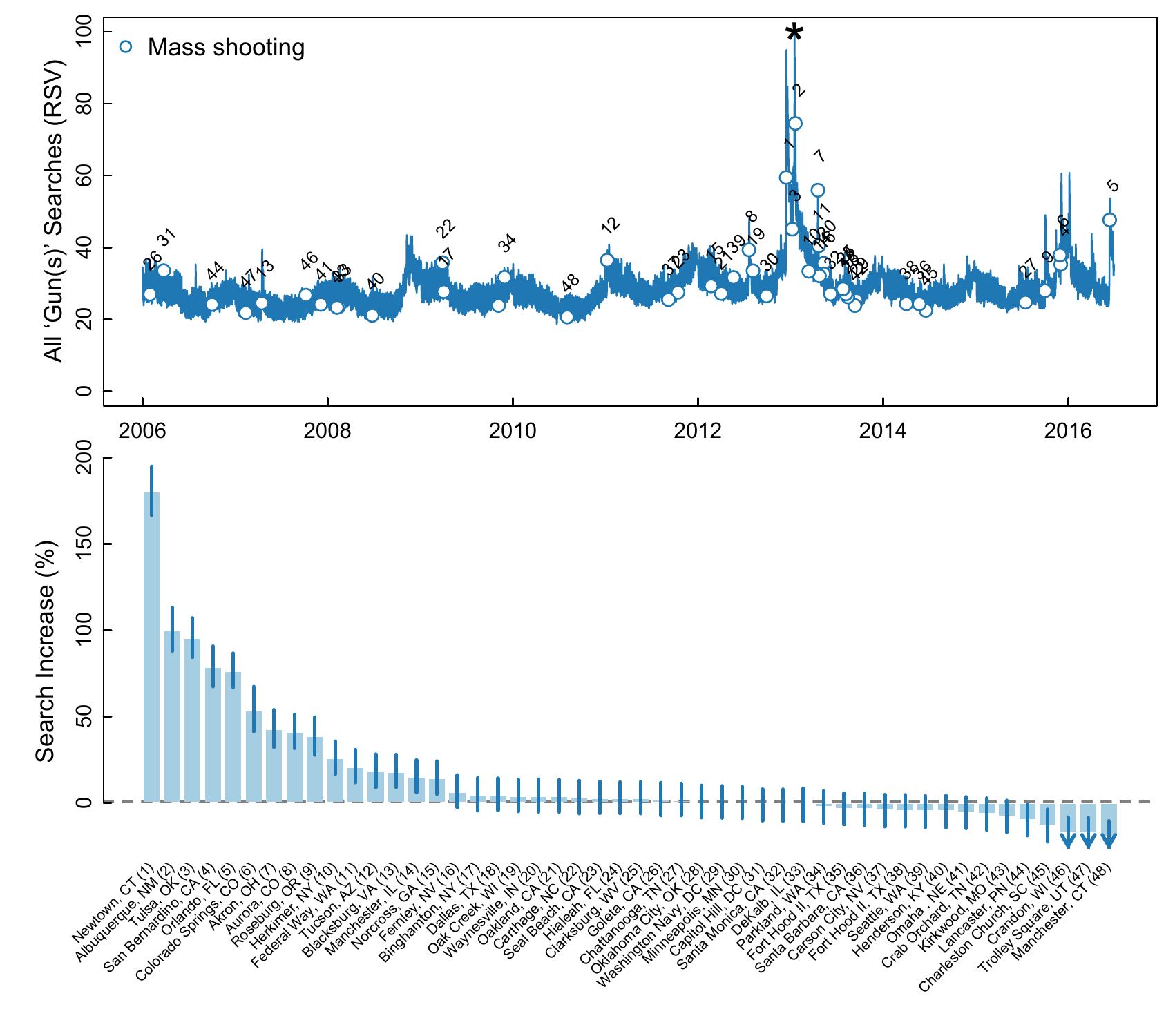}
\caption{Daily trends for all gun-related Google queries emerging from the United States (2006-2016).}
\label{fig:1}
\end{figure*}

\section{Missed Scientific Opportunities}
Many major improvements in the health and wellbeing of society derive from gathering epidemiological evidence about a disease and then applying the scientific method to test interventions that reduce harm or death from that disease. Just a brief sample of effective public health measures includes the widespread adoption of seatbelts in motor vehicles, the eradication of smallpox and routine vaccinations in general, the development of effective treatments for human immunodeficiency virus, and the prevention of smoking-attributable diseases through rigorous communication campaigns and other policy measures. Yet the power of science has essentially been locked out of one of the nation's most pressing health crises, the nation's seventh leading cause of mortality \cite{doi:10.1001/jama.291.10.1238} and the leading cause of death among persons aged 15 to 24: firearm injuries \cite{national2015health}.

Two decades ago, as gun violence prevention research was accelerating at the Center for Disease Control and Prevention's (CDC) National Injury Center, Congressman Jay Dickey (R-AR) amended the Omnibus Consolidated Appropriations Act of 1997 to read: ``None of the funds made available for injury prevention and control at the [CDC] may be used to advocate or promote gun control.'' \cite{104-208} In its wake, gun injury prevention funding within the CDC fell 96 percent to less than \$100,000 annually (a sum that cannot cover the costs of even a modest traditional study) \cite{guns2013access}. So under the thumb of the gun lobby was the CDC that the agency informed the National Rifle Association ``as a courtesy'' anytime investigators under their supervision studied issues related to gun violence, according to the New York Times \cite{nra_nytimes}. 

The Dickey Amendment was duplicative and unnecessary -- as a federal agency, the CDC is already prohibited from conducting advocacy -- and in a January 2013 memorandum \cite{obama_statement} President Obama clarified this and requested the CDC and other science agencies within Health and Human Services resume research on gun violence prevention \cite{legislativemandates}. But to date there is little new investment in gun violence research. The National Institutes of Health (the principal funder of public health research) has funded just two studies, and some health focused gun violence prevention research groups are relying on donations from the research staff themselves \cite{doi:10.1001/jama.2016.1707}.

Limitations have also been established on a range of firearm-related data. Beginning in 2003, Congress enacted a series of amendments restricting access to and use of crime gun trace data collected by the nation's law enforcement agencies and held by the Bureau of Alcohol, Tobacco, Firearms and Explosives. Those amendments continue to limit researchers' ability to analyze the movement of guns from the large group of lawful purchasers to the much smaller population who use them in crime.\footnote{\url{http://everytownresearch.org/reports/access-denied/}} Many states have also imposed restrictions, including exempting data on concealed carry permitting processes from public access. 

Although the scientific community agrees that gun violence prevention can be enhanced and informed by public health science \cite{doi:10.7326/0003-4819-158-9-201305070-00597,mozaffarian2013curbing,doi:10.1056/NEJMp1300512} -- even Congressman Dickey now supports gun violence prevention research -- without funding to collect the necessary data or explicit restrictions from accessing data, they cannot begin scientific inquiry to inform gun violence prevention. To provide some sense of the consequence of this gap in information, just 234 reports on {\em PubMed}, public health's publication database, include the term ``gun violence''\footnote{\url{http://www.ncbi.nlm.nih.gov/pubmed?term=\%22gun\%20violence\%22}} whereas, more than 88,000 studies indexed on PubMed make reference to ``influenza,''\footnote{\url{http://www.ncbi.nlm.nih.gov/pubmed/?term=influenza}} with gun violence rivaling influenza as a cause of premature death \cite{doi:10.1001/jama.291.10.1238}. The victims of gun violence cannot wait for legislation to unambiguously fund and support science that can inform prevention efforts. 

Fortunately, we believe that public big media data can yield new scientific insights into gun violence today.

\begin{figure}[t]
\centering
\includegraphics[width=.5\textwidth]{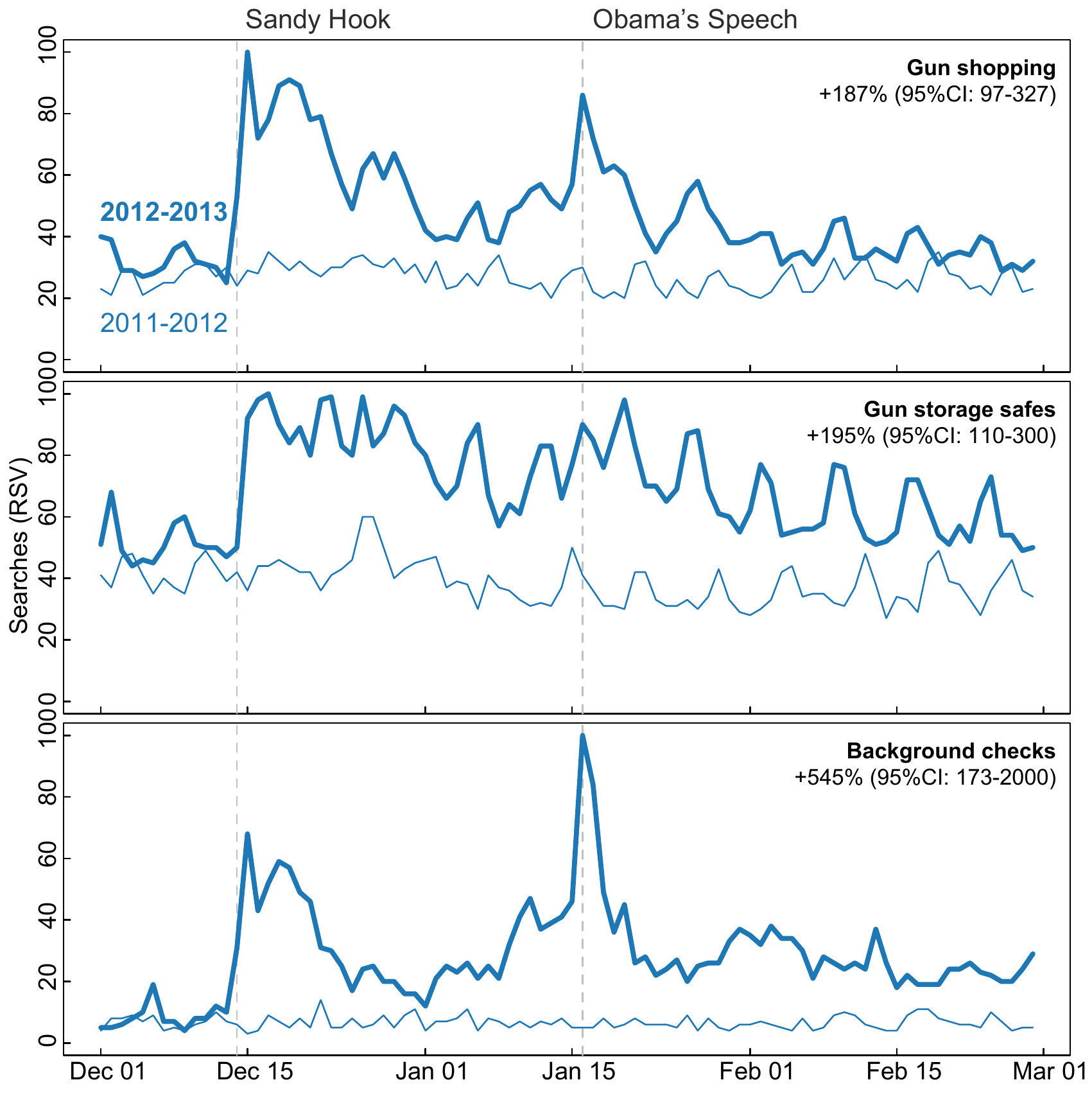}
\caption{Google searches for gun shopping, gun storage safes, and gun background checks increased after the Sandy Hook Elementary shooting.}
\label{fig:2}
\end{figure}

\section{Big Media Data Can Fill the Gap}

Big media data -- including large online media databases such as news aggregators, social media, and Internet searches -- are yielding significant breakthroughs across science, particularly in public health \cite{doi:10.1001/jama.2014.1505,Dredze:2012qy,doi:10.1056/NEJMp0900702,infodemiology}. In disease surveillance, researchers have developed a variety of new sources that accurately track \cite{culotta2010towards,signorini2011use} and forecast \cite{paul2014twitter,santillana2015combining,shaman2012forecasting} influenza by analyzing data from Twitter \cite{broniatowski2013national}, Web Searches \cite{ginsberg2009detecting,santillana}, specialized apps \cite{smolinski2015flu} and Wikipedia \cite{mciver2014wikipedia,hickmann2015forecasting}. These sources have also been used to rapidly respond to emerging infectious diseases, such as dengue fever \cite{althouse2011prediction} and  Zika \cite{bogoch2016anticipating,majumder2016utilizing,Dredze:2016oq}, and Ebola \cite{majumder20152014,carter2014twitter}. Big media data has yielded new insights into behavioral aspects of public health, including public responses to planned communication campaigns \cite{allem2016campaigns,ayers2015changes,ayers2015tips}, awareness campaigns, \cite{Ayers:2016rc,ayers2012novel,Owen01022006,Glynn2011} and organic/spontaneous events, such as a celebrity health disclosure \cite{Leas:2016qd,Ayers:2016fu,noar2015cancer,noar2013using,ayers2014celebrity}. Online resources are also helping to fight the current opioid epidemic \cite{katsuki2015establishing}, as well as studies that track emerging drugs \cite{lange2010salvia} and measure drug prevalence \cite{Michael-J.-Paul:2016fp}. These efforts are perhaps most valuable when they provide insights about phenomenon for which -- like gun violence -- there are  few existing high quality data resources. These include work on a broad array of mental health areas: measuring the prevalence of common mental illnesses like psychological distress \cite{Ayers:2012bj}; discovering seasonal patterns of mental illness \cite{ayers2013seasonality}, post traumatic stress disorder \cite{coppersmith2014measuring}, depression \cite{de2014characterizing,coppersmith2014quantifying}, and schizophrenia \cite{mitchell2015quantifying}; understanding eating disorder behaviors \cite{chancellor2016recovery}; and predicting suicidal ideation \cite{Choudhury:2016lq,Kumar:2015sf,jashinsky2014tracking,braithwaite2016validating,o2016talking}. That big media data provide insights into such a wide range of public health behaviors should give us hope that the same is possible for gun violence. 

A significant benefit of big media databases is that they are already free, publicly accessible, and timely, with users creating large volumes of text in real-time. For instance, there are more than 70 million annual Twitter postings that include terms like ``gun'' or ``guns,'' \cite{Benton:2016zh} that touch on the full spectrum of gun-related behaviors (e.g., ownership, safety practices, unlicensed sales, etc.) and gun-related attitudes (e.g., understanding of current gun safety measures, support for additional safety measures, etc.). Anecdotally, these data already influence the public's understanding of gun violence, as news reporting on many shootings rely on details gleaned from social media reports, like the Facebook Live broadcast of a police shooting in Minneapolis \cite{cnn_streaming}. Big media data could be a proverbial silver bullet to interject science into gun violence prevention, providing investigators the data with which to implement rigorous studies. 

A variety of private organizations are turning to strategies that generate, aggregate, and disseminate gun violence data to promote research. For instance, federal data on the occurrence of gun violence is limited and available only after significant delays. The nonprofit, nonpartisan Gun Violence Archive\footnote{\url{http://www.gunviolencearchive.org/}} aggregates incidents of gun violence from news reports, including both homicides and non-fatal shootings, and makes the data available in near real-time. 

But big media data scientists can go beyond simply counting incidents to answer far more detailed questions. For example, shootings are routinely covered in traditional news media and these contemporaneous records include many additional details about the circumstances such as the weapon and magazine used; the name, relationship, and prior criminal history of the shooter, and other details. The news article itself may also be processed to reveal the reporting frame: how the shooting is presented in ways beyond the factual details. Understanding media framing is important for understanding the way the public perceives gun safety issues \cite{mcginty2013effects}. For instance, Everytown for Gun Safety maintains a database of unintentional shootings involving children\footnote{\url{http://www.everytownresearch.org/notanaccident}}, which shows  that federal data vastly undercount the frequency of these tragic incidents, but more importantly the database shows that more than two-thirds of cases could have been prevented had the gun been stored responsibly. 

Another powerful application of these data can be demonstrated through an examination of Internet search queries for guns \cite{nuti2014use}. Using {\em Google Trends} \footnote{\url{http://www.google.com/trends}}, a public index of Google search volume, we analyzed all gun-related searches emerging from the United States that included the terms ``gun'' or ``guns'' and related search trends to fourty-eight of the most publicized of the more than one hundred thirty mass shoootings in the same period \footnote{\url{http://everytownresearch.org/mass-shootings/}} (Figure \ref{fig:1}). Gun-related searches have been on the rise recently, and appear to be spiking around America's spate of mass shootings since late 2012. For instance, gun-related searches reached record levels the day President Obama commented on the Sandy Hook Elementary shooting, and peaks also occurred in the days following the Sandy Hook Elementary shooting and notable mass shootings in San Bernardino and the Pulse nightclub in Orlando. 

Moreover, we can measure the behaviors and thoughts of the public towards guns via the content of their searches -- as we have done to describe the specific health concerns of the public in their own words -- \cite{althouse-population} and link these to specific strategies for gun violence prevention. Figure \ref{fig:2} shows Google searches for gun shopping (including all queries with the terms ``gun(s)'' and ``buy(s)'' or ``dealer(s)'' or ``shop(s)''), shopping for gun storage safes (``gun(s)'' and ``safe(s)''), and gun purchase background checks (``gun(s)'' and ``background(s)'') around the Sandy Hook Elementary shooting and President Obama's subsequent speech. Searches for all three domains significantly ($p < 0.05$) and substantially deviated from search volumes during the prior year. Searches for gun safe-related terms (e.g., ``buy a gun safe'' and like queries) increased 195\% (95\%CI: 110 to 300), spiking the day of the Sandy Hook Elementary shooting and remained elevated for two months. Whereas searches for gun shopping terms and gun background check terms spiked both on the day of the shooting and subsequently when President Obama publicly commented on gun violence and the spikes were less durable. 

\section{Future Directions}
As these case studies demonstrate, big media data have the potential to yield actionable insights for gun violence prevention research even without significant funding to support data collection, which is often the most costly component of the scientific process. Many unanswered questions about gun violence prevention can be explored this way, including those distinctly related to individual gun safety and public policy on gun safety.

Around individual gun safety the questions are only limited by the extent to which guns are discussed on big media channels. What other events trigger gun shopping? Is the public seeking out illicit gun paraphernalia such as after-market adapters that render weapons fully automatic? What safety aids (e.g., safes, trigger locks, etc.) are the public seeking? Unlike with traditional survey data, a big media data analysis of these questions has more face validity, where these behaviors are directly observed through Internet searches, social media posts, or microblogging. At the same time the structure of big media data, especially in social media, means the unit of analysis can be individuals (within established research ethics and privacy standards \cite{conway2014ethical}), meaning gun safety data can be linked to demographic traits (e.g., sex, ethnicity, education, etc.) that are associated with individual social media accounts. Moreover, investigators could study the co-occurrence of mitigating factors related to gun safety, such as mental illness and gun safety practices within individual social media account holders. 

At the policy level we can assess the public's understanding of existing and proposed gun safety regulation, summarize public feedback to inform the development of gun violence prevention regulations, and evaluate the extent to which regulations are being implemented or weakened by loopholes. For example, Everytown for Gun Safety tracked the emergence of a thriving online market for unlicensed gun sales without background checks,\footnote{\url{https://everytownresearch.org/reports/point-click-fire/}} which undermine existing gun safety laws by leaving an open door for convicted felons and domestic abusers to arrange gun sales and get armed with no questions asked. On a single website, they identified over 600,000 unique gun ads posted over a one-year-period, and the users' geographic location and other information listed publicly in the ads provides opportunities to study the frequency and character of this heretofore poorly understood commerce.\footnote{\url{https://everytownresearch.org/reports/business-as-usual}}

Effective gun safety regulation is dependent on an informed public. Public health has successfully invested in educating the public on effective regulatory strategies to promote safety (e.g., tobacco, seatbelts, underage drinking). In addition to assessing public understanding of gun safety using big media data, researchers can also use these same channels to disseminate messages that better inform the public \cite{mozaffarian2013curbing}.

Ultimately big media data provides a pathway to use the scientific method to broadly inform gun violence prevention by leveraging billions of diverse data points. It is in everyone's interest to have an empirically informed approach to gun violence prevention, with data driving us to the most effective strategies to improve public health. The speed of progress can be accelerated by data scientists and public health working together to harness big media data.

\bibliographystyle{abbrv}
\bibliography{refs}

\end{document}